\documentstyle[12pt,epsf]{article}
\begin{document}
\baselineskip 0.7cm

\newcommand{\gsim}{ \mathop{}_{\textstyle \sim}^{\textstyle >} }
\newcommand{\lsim}{ \mathop{}_{\textstyle \sim}^{\textstyle <} }
\newcommand{\vev}[1]{ \left\langle {#1} \right\rangle }
\newcommand{\lsp}{ \left ( }
\newcommand{\rsp}{ \right ) }
\newcommand{\lmp}{ \left \{ }
\newcommand{\rmp}{ \right \} }
\newcommand{\llp}{ \left [ }
\newcommand{\rlp}{ \right ] }
\newcommand{\labs}{ \left | }
\newcommand{\rabs}{ \right | }
\newcommand{\KEV}{ {\rm keV} }
\newcommand{\MEV}{ {\rm MeV} }
\newcommand{\GEV}{ {\rm GeV} }
\newcommand{\TEV}{ {\rm TeV} }
\newcommand{\mgut}{M_{GUT}}
\newcommand{\mint}{M_{I}}
\newcommand{\mgra}{M_{3/2}}
\newcommand{\mll}{m_{\tilde{l}L}^{2}}
\newcommand{\mdr}{m_{\tilde{d}R}^{2}}
\newcommand{\mllXX}[1]{m_{\tilde{l}L , {#1}}^{2}}
\newcommand{\mdrXX}[1]{m_{\tilde{d}R , {#1}}^{2}}
\newcommand{\mgy}{m_{G1}}
\newcommand{\mgl}{m_{G2}}
\newcommand{\mgc}{m_{G3}}
\newcommand{\nuR}{\nu_{R}}
\newcommand{\slL}{\tilde{l}_{L}}
\newcommand{\slLi}{\tilde{l}_{Li}}
\newcommand{\sdR}{\tilde{d}_{R}}
\newcommand{\sdRi}{\tilde{d}_{Ri}}
\newcommand{\e}[1]{{\rm e}^{#1}}

\begin{titlepage}

\begin{flushright}
TIT-HEP-307
\\
October, 1995
\end{flushright}

\vskip 0.35cm
\begin{center}
{\large \bf Suppression of Proton Decay \\in the Missing-Partner Model \\for
Supersymmetric SU(5) GUT\footnote
{
Talk given at Yukawa International Seminar '95: from the Standard
Model to Grand Unified Theories, Kyoto, Japan, August. 21-26. This talk is 
based on the work in collaboration with
T.~Moroi, K.~Tobe, and T.~Yanagida.
}
}
\vskip 1.2cm
J.~Hisano\footnote
{Fellow of the Japan Society for the Promotion of Science.}
\vskip 0.4cm

{\it Department of Physics, Tokyo Institute of Technology,\\ 
 Oh-okayama 2-12-1, Megro-ku, 152, Japan}

\vskip 1.5cm

\abstract{
The Peccei-Quinn symmetric extension of the missing-partner model in the supersymmetric
SU(5) grand unified model is consistent with the observed stability of the 
proton, even in the large
$\tan\beta$ region ($\simeq 50-60$) expected from the Yukawa unification. 
Moreover, the SU(5) gauge coupling constant remains small enough for the perturbative 
description of GUT's below the gravitational scale. 
}

\end{center}
\end{titlepage}

%
%
%
%

\section{Introduction}
	Supersymmetry (SUSY) is introduced to protect the weak scale from the 
radiative corrections \cite{ACTB12-437}. Since supersymmetry is a symmetry between bosons
and fermions, the minimal supersymmetric extension of standard model (MSSM) 
introduces scalar bosons with baryon or lepton numbers, which are called as  squarks 
or sleptons.  Therefore, there may be baryon or lepton number violating 
operators with the dimensions four or five, which do not exit in 
the standard model. The dimension-four operators induce various 
phenomenological difficulties, especially, instability of proton.
Therefore, they are expected to be forbidden by a symmetry, and a $Z_2$ 
symmetry called as R parity is often assumed phenomenologically.

Quark-quark-squark-slepton is a dimension-five operator violating both baryon
and lepton numbers. This is suppressed by a mass parameter, which we call 
as $\Lambda$ from now on. The magnitude of $\Lambda$ depends on the origin 
of the dimension-five operator. In the supergravity model $\Lambda$ is 
expected to be $m_{pl}/\sqrt{8\pi}$$(=2.4\times10^{18}$GeV), and then 
the proton lifetime 
is $10^{(26-28)}$ years. This value is lower than the experimental
lowerbound by the magnitude of $10^{-(4-6)}$.  The dimension-five operator 
is also expected to be forbidden by a symmetry.

One of the candidates for such a symmetry 
is the Peccei-Quinn (PQ) symmetry \cite{PRL38-1440}, which is a solution of
strong CP problem.  This is because a U(1) symmetry 
forbidding the dimension-five operator must have a triangle anomaly. 
On the other hand, the breaking scale of the PQ symmetry ($M_{PQ}$) is 
constrained from astronomy and cosmology as following \cite{kim},
\begin{equation}
10^{10} {\rm GeV} \le M_{PQ} \le 10^{13} {\rm GeV}.
\label{pqscale}
\end{equation}
Therefore, the PQ symmetry can not forbid the dimension-five operator
completely, however, can suppress it by a factor of $(M_{PQ}/\Lambda)$. 
If $\Lambda$ is $2.4\times10^{18}$ GeV, the 
proton lifetime can reach $10^{(36-46)}$ years, and we can avoid the constraint from 
the observation. The PQ symmetry is expected not only from the strong
CP problem, but also from the stability of proton.

Next, we will extend the SU(5) supersymmetric grand unified theory (SU(5)
SUSY GUT)\cite{NPB193-150}, 
which is very interesting from both experimental and theoretical points of 
view, to have the PQ
symmetry. In the minimal SU(5) SUSY GUT the dimension-five operator comes 
from Yukawa couplings giving masses to quarks and leptons. Quarks and
leptons are embedded in $\psi_i$[{\bf 10}] and $\phi_i$[{\bf 5}$^\star$]
($i$=1,2,3).
A pair of SU(2)$_L$-doublet Higgses in MSSM, $H_f$ and $\overline{H}_f$,
is in $H$[{\bf 5}] and $\overline{H}$[{\bf 5$^*$}] as
\begin{eqnarray}
H^A &=& \left(
\begin{array}{ccccc}
H_c, & H_c, & H_c, & H_f, & H_f
\end{array}
\right)^t,
\nonumber\\
\overline{H}_A &=& \left(
\begin{array}{ccccc}
\overline{H}_c, & \overline{H}_c, &\overline{H}_c, 
& \overline{H}_f, & \overline{H}_f
\end{array}
\right)^t,
\label{h5}
\end{eqnarray}
where $A(=1\cdots5)$ is a SU(5) fundamental representation index. 
Here, two color-triplet Higgses, $H_c$ and $\overline{H}_c$, have to be 
introduced in $H$ and $\overline{H}$. The SU(5) symmetric Yukawa couplings
are given as 
\begin{equation}
W_{\rm Yukawa} = 
\frac14 f_u^{ij} \epsilon_{ABCDE} \psi_i^{(AB)}\psi_j^{(CD)} H^E
+\sqrt{2} f_d^{ij} \psi_i^{(AB)}\phi_{jA} \overline{H}_B,
\label{yukawa}
\end{equation}
where $\epsilon_{ABCDE}$ is a fifth antisymmetric tensor.
The dimension-five operator is generated by an exchange of the color-triplet Higgses through these Yukawa couplings
\cite{NPB197-533}. 
The present lower bound of the color-triplet Higgs mass from the negative search 
of proton decay has already reached at $2\times10^{16}$ GeV\cite{nath,NPB402-46,tu-490}, and the minimal 
SU(5) SUSY GUT is also strongly constrained now.

If the minimal SU(5) SUSY GUT is extended to have the PQ symmetry, there 
may be no problem of the dimension-five operator. However, the color-triplet Higgs 
mass must vanish, and proton can decay very rapidly again through the 
dimension-six operator induced by an exchange of color-triplet Higgs boson. In order to 
preserve the PQ symmetry in the Yukawa couplings (\ref{yukawa}), each chiral 
multiplets are transformed under the PQ symmetry as 
\begin{eqnarray}
\psi_i  &\rightarrow& \e{i\alpha} \psi_i, 
\nonumber\\
\phi_i  &\rightarrow& \e{i\beta} \phi_i, 
\nonumber\\
H    &\rightarrow& \e{-2 i\alpha} H 
,\nonumber\\
\overline{H}    &\rightarrow& \e{-i(\alpha+\beta)} \overline{H},
\label{PQ-charge0}
\end{eqnarray} 
where $3\alpha + \beta \ne 0$ to forbid the dimension-five operator. 
Therefore, $H$ and $\overline{H}$ can not 
have an SU(5) symmetric mass term, and the color-triplet Higgs masses 
must be zero.  

A simple way for the color-triplet Higgses to have a GUT-scale mass is to introduce 
new {\bf 5}- and {\bf 5}$^*$-dimension Higgses ($H^\prime$ and 
$\overline{H}^\prime$) with the ${\rm U(1)}_{PQ}$ charges opposite to 
$\overline{H}$ and $H$\cite{PLB291-263}. However, this extension 
violates a experimental success 
of gauge coupling unification. In this extension there is another pair of 
SU(2)$_L$-doublet Higgses in $H^\prime$ and $\overline{H}^\prime$, and these 
can acquire a mass only from the vacuum expectation value breaking the PQ
symmetry. These extra SU(2)$_L$-doublet Higgses give
extra corrections to the SU(2)$_L$ and U(1)$_Y$ gauge coupling constants.
Therefore, this extension breaks the successful gauge coupling unification. 
To reproduce the gauge coupling unification, we have to
introduce many particles with the masses at the GUT scale 
so that the threshold 
corrections to the three gauge coupling constants can
compensate those from these extra SU(2)$_L$-doublet Higgses.

In next section, we will propose the missing-partner model with the PQ
symmetry. Since this model gives the large threshold corrections to the three 
gauge coupling constants at the GUT scale generically \cite{PRL70-709}, 
the success of gauge coupling unification can be acquired naturally, and 
the PQ symmetry can suppress  sufficiently the proton decay through the
dimension-five operator\cite{PLB324-138}.
 
\section{The Missing-Partner Model with the Peccei-Quinn symmetry}

The missing-partner model was proposed for the SU(2)$_L$-doublet Higgses in 
MSSM not to receive, group-theoretically, a mass from the 
SU(5)-breaking vacuum expectation
value\cite{PLB115-380}. This model has three Higgses, $\Sigma$[{\bf 75}], 
$\theta$[{\bf 50}], and $\overline{\theta}$[{\bf 50$^\star$}] with 
$H$ and $\overline{H}$ that we have mentioned above. The superpotential
has following terms,
\begin{eqnarray}
\label{superpotential1}
W &=& 
  G_H H^A \Sigma_{(FG)}^{(BC)} \theta^{(DE)(FG)} \epsilon_{ABCDE} 
+ G_{\overline{H}} \overline{H}_A \Sigma^{(FG)}_{(BC)}\overline{\theta}_{(DE)(FG)} \epsilon^{ABCDE} 
\nonumber\\ &&
+M_{75}\Sigma_{(CD)}^{(AB)}\Sigma_{(AB)}^{(CD)}
-\frac13 \lambda_{75} \Sigma_{(EF)}^{(AB)}\Sigma_{(AB)}^{(CD)}\Sigma_{(CD)}^{(EF)}
\nonumber\\ &&
+W_{\rm Yukawa}
\end{eqnarray}
with a mass term of $\theta$ and $\overline{\theta}$ 
($M_{50}\overline{\theta}_{(AB)(CD)}\theta^{(AB)(CD)}$).
The {\bf 75}-dimension
Higgs $\Sigma$ acquires a vacuum expectation value and it breaks 
the SU(5) symmetry. 
There can be no $H \Sigma \overline{H}$ term due to the SU(5) gauge symmetry,
and the {\bf 50}-dimension Higgses, $\theta$ and $\overline{\theta}$,
have SU(3)$_C$ triplet components, however, no SU(2)$_L$-doublet component. 
Therefore, $\langle \Sigma \rangle$ can give masses to the color-triplet 
Higgses, however, not to the SU(2)$_L$-doublet Higgses.

The missing-partner model has two different behaviors of the running gauge 
coupling constants from the minimal SU(5) SUSY GUT since this model has large 
dimension Higgses. First, the SU(5) gauge coupling constant blows up 
rapidly above the mass of $\theta$ and $\overline{\theta}$, $M_{50}$. If 
a perturbative picture is expected not to be broken below the gravitational 
scale, $M_{50}$ should be at least above the gravitational scale. However, in that case the 
color-triplet Higgses inducing proton decay have a smaller mass than the GUT
scale since they have a see-saw type mass matrix. In a case where 
$M_{50}$ is $2.4\times10^{18}$GeV, the color-triplet Higgs mass becomes $10^{(14-15)}$ 
GeV, that is completely excluded experimentally.

Second, the {\bf 75}-dimension Higgs $\Sigma$ gives large threshold 
corrections to the SU(3)$_C$\\$\times$SU(2)$_L$$\times$U(1)$_Y$ gauge coupling 
constants. The components of $\Sigma$ acquire different masses from 
each others due to its own vacuum expectation value as following 
table~\cite{PRL70-709}. 
~~\\~~\\
\begin{center}
\begin{tabular}{cc} \hline \hline
$({\rm SU(3)}_C\times{\rm SU(2)}_L\times{\rm U(1)}_Y)$  &  mass \\ \hline
$({\bf 8},{\bf 3}, { 0})$ & $M_{\Sigma}$ \\
$({\bf 3},{\bf 1}, {\frac53})$,
$({\bf \overline{3}},{\bf 1},{ -\frac53})$ & $\frac45 M_{\Sigma}$\\
$({\bf 6},{\bf 2},{\frac56})$,
$({\bf \overline{6}},{\bf 2},{-\frac56})$ & $\frac25 M_{\Sigma}$ \\
$({\bf 1},{\bf 1} ,{0})$ & $\frac25 M_{\Sigma}$ \\
$({\bf 8},{\bf 1}, {0})$ & $\frac15 M_{\Sigma}$ \\
$({\bf 3},{\bf 2}, {-\frac56})$,
$({\bf {\overline{3}}},{\bf 2}, {\frac56})$ & {\rm
Nambu-Goldstone multiplets}\\ \hline
\end{tabular}
\end{center}
~\\~\\
Here $M_{\Sigma}=5M_{75}$.
This mass splitting contributes to the differences
of the three gauge coupling constants. This was needed surely when we
would extend the minimal SU(5) SUSY GUT to have the PQ symmetry.

The missing-partner model has a more severe problem for the
stability of proton if we assume the perturbative picture bellow the
gravitational scale. However, since there are large threshold 
corrections to the gauge coupling constants at the GUT scale, the problem 
is expected to be solved completely by extending this model to have 
the PQ symmetry.

To preserve the PQ symmetry in the superpotential 
(\ref{superpotential1}), $\Sigma$, $\theta$, and $\overline{\theta}$ 
are transformed as
\begin{eqnarray}
\theta({\bf 50}) &\rightarrow& e^{2 i\alpha} \theta({\bf 50}),
\nonumber \\
\overline{\theta}({\bf \overline{50}}) &\rightarrow& e^{i(\alpha+\beta)} 
\overline{\theta}({\bf \overline{50}}),
\nonumber \\
\Sigma({\bf 75}) &\rightarrow& \Sigma({\bf 75}),
\label{PQ-charge1}
\end{eqnarray}
with the other chiral multiplets transformed as Eqs.~(\ref{PQ-charge0}).
Here, we have to introduce newly $H^\prime$[{\bf 5}], 
$\overline{H}^\prime$[{\bf 5}$^\star$], $\theta^\prime$[{\bf 50}], 
and $\overline{\theta}^\prime$[{\bf 50}$^\star$] with the  ${\rm U(1)}_{PQ}$ charges
opposite to the corresponding chiral multiplets. This is 
also because these Higgses have the PQ symmetric masses. 
We add a new superpotential to Eq.~(\ref{superpotential1}),
\begin{eqnarray}
\label{superpotential2}
W^{'} &=& 
 G_H^{'} H^{\prime A} \Sigma_{(FG)}^{(BC)} \theta^{\prime(DE)(FG)} \epsilon_{ABCDE} 
+G_{\overline{H}}^{\prime} \overline{H}_A^{\prime} \Sigma^{(FG)}_{(BC)}
\overline{\theta}_{(DE)(FG)}^{\prime} \epsilon^{ABCDE} 
\nonumber \\ &&
 + M_1 \overline{\theta}_{(AB)(CD)}^{'} \theta^{(AB)(CD)}+ M_2 \overline{\theta}_{(AB)(CD)} \theta^{\prime(AB)(CD)}.
\end{eqnarray}
To avoid that the SU(5) gauge coupling constant blows up below the
gravitational scale $M_{pl}/\sqrt{8\pi}$, we assume 
\begin{eqnarray}
M_1,~M_2 { \mathop{}_{\textstyle \sim}^{\textstyle >} } 10^{18}{\rm GeV} .
\end{eqnarray}
In the following, we take $M_1=M_2$$=M_{pl}/\sqrt{8\pi}$($\equiv 2.4\times10^{18}$GeV) for simplicity. Then, we have four Higgses, $H$, $\overline{H}$, $H^{'}$,
$\overline{H}^{'}$, and one Higgs $\Sigma$ much below the
gravitational scale.

The {\bf 75}-dimension Higgs $\Sigma$
has the following vacuum expectation value that causes the breaking
SU(5) $\rightarrow$ 
SU(3)$_C$$\times$SU(2)$_L$$\times$U(1)$_Y$,
\begin{eqnarray}
\langle \Sigma \rangle_{(\gamma\delta)}^{(\alpha\beta)} 
&=& \frac12 \left\{
\delta_\gamma^\alpha \delta_\delta^\beta -
\delta_\delta^\alpha \delta_\gamma^\beta
\right\} V_{\Sigma},
\nonumber\\
\langle \Sigma \rangle_{(cd)}^{(ab)} 
&=& \frac32 \left\{
\delta_c^a \delta_d^b -
\delta_d^a \delta_c^b
\right\} V_{\Sigma},
\\
\langle \Sigma \rangle_{(b \beta)}^{(a \alpha)} 
&=& -\frac12 \left\{
\delta_b^a \delta_\beta^\alpha
\right\} V_{\Sigma},
\nonumber
\end{eqnarray}
where
\begin{equation}
V_{\Sigma}=\frac32 \frac{M_{75}}{\lambda_{75}}.
\end{equation}
Here, $\alpha, \beta\dots$ are the SU(3)$_C$ indices and $a, b\dots$ the
SU(2)$_L$ indices. This vacuum expectation value generates masses for the
color-triplet Higgses as (after integrating out the heavy fields, $\theta$,
$\overline{\theta}^\prime$ and $\theta^\prime$, $\overline{\theta}$),
\begin{eqnarray}
  M_{H_c} H_c^\alpha \overline{H}_{c\alpha}^{\prime} 
+ M_{\overline{H}_c} H_c^{\prime \alpha} \overline{H}_{c\alpha},
\end{eqnarray}
with
\begin{eqnarray}
\label{coloredmasses}
M_{H_c} \simeq 48 G_H G_{\overline{H}}^{'} \frac{V_{\Sigma}^2}{M_1},~~~
M_{\overline{H}_c} \simeq 48 G_{\overline{H}}G_H^{'} \frac{V_{\Sigma}^2}{M_2}.
\end{eqnarray}
For $V_\Sigma\simeq 10^{(15-16)}$GeV and $G_H G_{\overline{H}}^\prime
\sim G_{H}^\prime G_{\overline{H}} \sim 1$, $M_{H_c}$
and  $M_{\overline{H}_c}$ are at $10^{(14-15)}$GeV. The four 
${\rm SU(2)}_L$-doublet Higgses, $H_f$, $\overline{H}_f$,
$H_f^{\prime}$, and $\overline{H}_f^{\prime}$, remain massless.

In order to break the PQ symmetry, we introduce a SU(5)-singlet 
chiral multiplet $P$  whose ${\rm U(1)}_{PQ}$ charge is chosen
as $P\rightarrow e^{-i(3\alpha+\beta)}P$ so that the following
superpotential is allowed,
\begin{eqnarray}
\label{PQpotential}
W^{''} = g_P \overline{H}_A^{\prime} H^{\prime A} P.
\label{pq-yukawa}
\end{eqnarray}
The vacuum expectation value of $P$, which is constrained to be at $M_{PQ}$
(see Eq.~(\ref{pqscale})), gives
an intermediate-scale mass to a pair of SU(2)$_L$-doublet Higgses, 
$H_f^{\prime}$ and $\overline{H}_f^{\prime}$, 
\begin{eqnarray}
M_{H_f^{'}} = g_P {\langle P \rangle}.
\end{eqnarray}
The mechanism of breaking the PQ symmetry  at the intermediate scale
will be discussed in next section.

The color-triplet Higgses have an off-diagonal element 
in their mass matrix as
\begin{eqnarray}
\left( \overline{H}_c, \overline{H}_c^{\prime} \right)
\left(
\begin{array} {cc}
M_{\overline{H}_c} & 0 \\
g_P\langle P  \rangle     &M_{H_c}
\end{array}
\right)
\left(
\begin{array} {c}
H_c^{\prime}\\
H_c
\end{array}
\right).
\end{eqnarray}
The baryon-number violating dimension-five operator
mediated by the color-triplet
Higgses is given in the present model as (see Fig.~1)
\begin{eqnarray}
\label{missingD=5}
\frac{g_P\langle P \rangle }{M_{H_c} M_{\overline{H}_c}} \frac{1}{2\sqrt{2}} 
f^{ij}_d f^{kl}_u
\left( \phi_{F i} \psi_j^{(FA)} \right) 
\left( \psi_k^{(BC)} \psi_l^{(DE)} \right)
 \epsilon_{ABCDE}.
\end{eqnarray}
        \begin{figure}
           \epsfxsize= 6 cm   
           \centerline{\epsffile{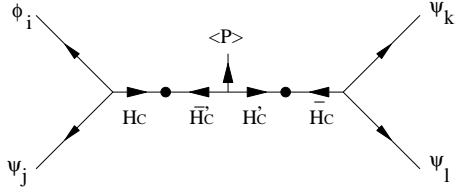}}
        \caption{The Feynman diagram of the baryon-number violating 
                 dimension-five operator in the present model.}
        \label{fig:1}
        \end{figure}
Notice that the pre-factor of the dimension-five operator in the original 
missing-partner model is ${f^{ij}_d f^{kl}_u}/{2\sqrt{2} M_{H_c}}$.
Thus, we easily see that the dimension-five operator in the present model is
more suppressed by a factor $M_{H_f^{\prime}}/M_{H_c}$, and 
\begin{equation}
\frac{M_{H_c} M_{\overline{H}_c}}{M_{H_f^{\prime}}}
\simeq 10^{(18-20)} {\rm GeV},
\label{dimen5}
\end{equation}
if  $M_{H_f^{'}}=10^{10}$GeV and  $M_{H_c}
\sim M_{\overline{H}_c}\sim 10^{(14-15)}$GeV. This value is 
consistent with the present negative observation of proton decay\cite{nath,NPB402-46,tu-490},  even if
$\tan\beta(\equiv \langle H_f \rangle/\langle \overline{H}_f \rangle) 
=(50-60)$ in which region the intriguing idea of the Yukawa
coupling unification $f^t=f^b=f^\tau$ at the GUT scale
\cite{yukawa-u} is consistent.

Now we will point out that this extension is consistent with the gauge 
coupling unification. The mass spectrum above the PQ symmetry breaking scale 
contains four SU(2)$_L$-doublets Higgses,  and the success of the gauge 
coupling unification seems to be lost at first sight. However, 
$\Sigma$({\bf75}) gives large threshold corrections to the three 
gauge coupling constants as we mentioned before, and the corrections  
can compensate those from the extra SU(2)$_L$-doublet Higgses.
The running of the three gauge coupling constants 
at the one-loop level is given by the following solutions to the 
renormalization group equations,
\begin{eqnarray}
\alpha_3^{-1} (m_Z) &=& \alpha_{5}^{-1} (\Lambda)
+ \frac{1}{2\pi} \Bigg\{ 
\left( -2 - \frac{2}{3} N_g \right) \ln \frac{m_{SUSY}}{m_Z}
+ (-9 + 2 N_g) \ln \frac{\Lambda}{m_Z}
  \nonumber \\
&&
-4 \ln \frac{\Lambda}{M_V}  
 + \ln \frac{\Lambda}{M_{H_c}}
 + \ln \frac{\Lambda}{M_{\overline{H}_c}} 
\nonumber\\
&&
        + 9 \ln \frac{\Lambda}{M_\Sigma} 
        +   \ln \frac{\Lambda}{0.8 M_\Sigma}
        + 10\ln \frac{\Lambda}{0.4 M_\Sigma}
	+ 3 \ln \frac{\Lambda}{0.2 M_\Sigma}
\Bigg\}, 
\label{alpha3}
\\
\alpha_2^{-1} (m_Z) &=& \alpha_{5}^{-1} (\Lambda)
+ \frac{1}{2\pi} \Bigg\{ 
\left( -\frac{4}{3} - \frac{2}{3} N_g - \frac{5}{6} \right)\ln \frac{m_{SUSY}}{m_Z}
+ (-6 + 2 N_g + 1) \ln \frac{\Lambda}{m_Z}
  \nonumber \\
&&
-6 \ln \frac{\Lambda}{M_V} 
+ \ln \frac{\Lambda}{M_{H_f^\prime}}
\nonumber\\
&&
	+ 16 \ln \frac{\Lambda}{M_\Sigma}
	+ 6  \ln \frac{\Lambda}{0.4 M_\Sigma}
		\Bigg\}, 
\label{alpha2}
\\
\alpha_1^{-1} (m_Z) &=& \alpha_{5}^{-1} (\Lambda)
+ \frac{1}{2\pi} \Bigg\{ 
\left( -\frac{2}{3} N_g - \frac{1}{2} \right)\ln \frac{m_{SUSY}}{m_Z}
 + \left(2 N_g + \frac{3}{5} \right) \ln \frac{\Lambda}{m_Z}
 \nonumber \\
&&
-10 \ln \frac{\Lambda}{M_V} 
+ \frac25 \ln \frac{\Lambda}{M_{H_c}}
+ \frac25 \ln \frac{\Lambda}{M_{\overline{H}_c}} 
+ \frac35 \ln \frac{\Lambda}{M_{{H}_f^\prime}}
\nonumber\\
&&
	+ 10 \ln \frac{\Lambda}{0.8 M_\Sigma}
	+ 10 \ln \frac{\Lambda}{0.4 M_\Sigma} 
 \Bigg\},
\label{alpha1}
\end{eqnarray}
where $\alpha_5 \equiv g^2_5 /4\pi$ is the SU(5) gauge coupling
constant, $M_V$ the heavy gauge boson mass ($M_V=2\sqrt{15}g_5
V_{\Sigma}$), and $\Lambda$ the renormalization point which is taken
to be much larger than the GUT scale.
Here, we have assumed that all superparticles in MSSM
 have a SUSY-breaking common mass $m_{SUSY}$ for simplicity, and 
the mass splitting of $\Sigma$({\bf 75}) has been included.
 By eliminating
$\alpha^{-1}_{5}$ from Eqs.~(\ref{alpha3}-\ref{alpha1}), we obtain
simple relations \cite{NPB402-46,PLB291-263,PLB324-138}:
\begin{eqnarray}
\label{MHC}
(3 \alpha_2^{-1} - 2 \alpha_3^{-1} - \alpha_1^{-1}) (m_Z)
	&=& \frac{1}{2\pi} \Bigg\{ 
		\frac{12}{5} \, \ln \frac{M_{H_c}M_{\overline{H}_c}}{M_{H_f^\prime}m_Z}
		- 2 \, \ln \frac{m_{SUSY}}{m_Z} \nonumber\\
&&	         - \frac{12}{5} \, \ln (1.7\times 10^4)
\Bigg\},\\
\label{MVMSIGMA}
(5 \alpha_1^{-1} - 3 \alpha_2^{-1} - 2 \alpha_3^{-1}) (m_Z)
	&=& \frac{1}{2\pi} \Bigg\{
		12 \, \ln \frac{M_V^2 M_\Sigma}{m_Z^3}
		+ 8 \ln \frac{m_{SUSY}}{m_Z}  \nonumber\\ 
&& 		 + 36 \, \ln (1.4)
\Bigg\}.
\end{eqnarray}
Notice that the last terms in Eqs.~(\ref{MHC},\ref{MVMSIGMA}) come
from the mass splitting of $\Sigma$({\bf 75}), which makes a crucial
difference between the present model and the extension of the minimal
SU(5) SUSY GUT where {\bf 24}-dimension Higgs breaks the SU(5) symmetry.

To perform a quantitative analysis, we use the two-loop renormalization
group equations between the weak and the GUT scales.  Instead of the
common mass $m_{SUSY}$ of superparticles we have used the mass spectrum
estimated from the minimum supergravity \cite{NPB402-46,nojiri} to calculate the one-loop
threshold correction at the SUSY-breaking scale. Using the experimental
data $\alpha^{-1}(m_Z)=127.9\pm 0.1$, $\sin^2\theta_W(m_Z)=0.2315\pm
0.0003$, and $\alpha_3(m_Z)=0.116\pm 0.005$~\cite{coupling_data},
we obtain 
\begin{eqnarray}
\label{missingMHC}
1.9 \times 10^{17}~{\rm GeV} \leq 
&\frac{M_{H_c}M_{\overline{H}_c}}{M_{H_f^\prime}}& 
\leq 1.3 \times 10^{20}~{\rm GeV},\\
\label{missingMGUT}
9.1 \times 10^{15}~{\rm GeV} \leq 
&(M_V^2 M_\Sigma)^{1/3}& 
\leq 1.7 \times 10^{16}~{\rm GeV}.
\end{eqnarray}
The value of Eq.~(\ref{missingMHC}) is very much consistent 
with Eq.~(\ref{dimen5}). Notice that this  reason comes from the presence 
of the constant term in Eq.~(\ref{MHC}) which originates from the mass
splitting of $\Sigma({\bf 75})$.  

We show the evolution of the
SU(3)$_C$$\times$SU(2)$_L$$\times$U(1)$_Y$ and the 
SU(5) gauge coupling constants in Fig.~2 taking
$M_{H_c}=M_{\overline{H}_c}=10^{15}$GeV and $M_{H_f^\prime}=10^{10}$GeV for a
demonstrational purpose. The unification of the three gauge
coupling constants occurs around $10^{16}$GeV and the SU(5) gauge
coupling constant stays in the perturbative regime below the
gravitational scale, 2.4$\times 10^{18}$GeV.
        \begin{figure}
           \epsfysize= 6 cm   
           \centerline{\epsffile{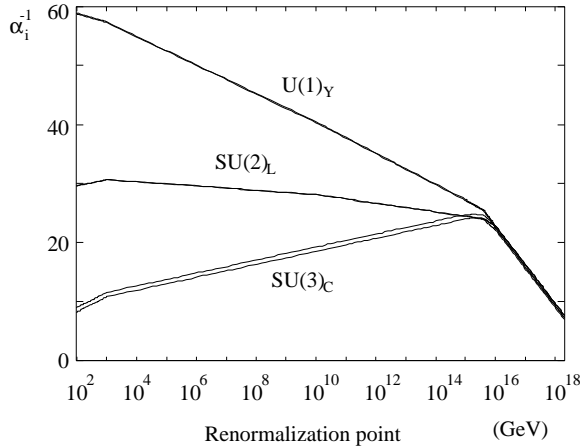}}
        \caption{ The flows of the running gauge coupling constants of
SU(3)$_C$ $\times$SU(2)$_L$$\times$U(1)$_Y$ and SU(5). Here, $M_{H_c}$
and $M_{\overline{H}_c}$ are taken at $10^{15}$GeV, and $M_{H_f^\prime}$
at $10^{10}$GeV.  We assume the SUSY-breaking scale $\sim$1TeV.}
        \label{fig:2}
        \end{figure}
\section{Conclusions and Discussion}
The Peccei-Quinn symmetric extension of the missing-partner model in SUSY SU(5) GUT 
is consistent with the observed stability of the proton, even if the masses 
of the unwanted {\bf 50}-dimension Higgses are lifted up to the gravitational
scale so that the SU(5) gauge coupling constant remains small enough for the 
perturbative description of GUT's. Moreover, even the large $\tan\beta$ 
region, expected from the Yukawa unification, is allowed from the 
experimental constraint. Here, we will comment some points for this model.

First, so far we have assumed that $P$ acquires the vacuum expectation value at
$M_{PQ}$ without the explicit potential. This is possible if we introduce another
SU(5)-singlet chiral multiplet $Q$, whose  ${\rm U(1)}_{PQ}$ charge is chosen
as $Q\rightarrow e^{3 i(3\alpha+\beta)}Q$ so that the following
superpotential is allowed~\cite{PLB291-418},
\begin{eqnarray}
W^{'''} = \frac{f}{M} P^3 Q.
\end{eqnarray}
These Higgses, $P$ and $Q$, have a very flat scalar potential  as
\begin{eqnarray}
V(P,Q) = \frac{f^2}{M^2} |P|^6 + \frac{f^2}{M^2} |3P^2Q|^2.
\end{eqnarray}
If a negative soft SUSY breaking mass $\sim - m^2$ for $P$ is introduced 
by any strong Yukawa coupling, it can induces very naturally the 
PQ symmetry breaking \cite{PLB291-418}, 
\begin{eqnarray}
\langle{P}\rangle \simeq \langle{Q}\rangle 
\simeq \sqrt{\frac{M m}{f}} \sim 10^{11} {\rm GeV},
\end{eqnarray}
provided $m \sim 1$TeV and $f\sim 1$.

Next, two independent charges $\alpha$ and $\beta$ defined in 
Eqs.~(\ref{PQ-charge0},\ref{PQ-charge1}) mean the presence of 
two global U(1) symmetries. 
If we introduce right-handed neutrino multiplets $N_i$ ({\bf 1}) 
and add the following terms to the superpotential,
\begin{eqnarray}
\label{majorana}
W^{''''} = k_{ij} N_i \phi_j H + j_{ij} N_i  N_j P,
\end{eqnarray}
we have only one ${\rm U(1)}$ symmetry and the charge $\alpha$ 
is fixed as $\alpha = 3\beta$\cite{MOD-A1-541}. The Yukawa couplings 
$j_{ij}N_i N_j P$ 
in Eq.~(\ref{majorana}) induce Majorana masses for the right-handed 
neutrino multiplets $N_i$, with $\langle P\rangle \ne 0$. It is 
interesting that the Majorana masses for the right-handed 
neutrinos are expected to be $O(10^{11})$ GeV, which naturally induce
very small masses of neutrinos through the cerebrated see-saw
mechanism~\cite{seesaw1} in a range of the MSW solution~\cite{nc9c-17} 
to the solar neutrino problem.

\section*{Acknowledgements}
 The author would like to thank K.~Tobe for careful reading this
manuscript.
%
%
\newcommand{\PL}{Phys.~Lett.}
\newcommand{\PR}{Phys.~Rev.}
\newcommand{\PRL}{Phys.~Rev.~Lett.}
\newcommand{\NP}{Nucl.~Phys.}
\newcommand{\ZP}{Z.~Phys.}
\newcommand{\PTP}{Prog.~Theor.~Phys.}
\newcommand{\NC}{Nuovo~Cimento}

\end{document}